\documentclass{cool}
\usepackage{amsmath,amssymb,amsfonts}

\newcommand{\lambdabar}{{\hbox{$\lambda_e$\kern-1.9ex\raise+0.45ex\hbox{--}
\kern+0.2ex}}}

\title{Accelerator Cavities as a Probe of Millicharged Particles}
\shorttitle{Accelerator Cavities as a Probe \ldots}
\author{H. Gies\inst{1} \and J. Jaeckel\inst{2} \and A.
  Ringwald\inst{2}}

\institute{                    
  \inst{1} Institut f\"ur Theoretische Physik, Universit\"at Heidelberg, Philosophenweg 16, 
D-69120 Heidelberg, Germany\\ 
  \inst{2} Deutsches Elektronen-Synchrotron DESY, Notkestra\ss e 85, D-22607 Hamburg, Germany
}


\begin{document}

\maketitle

\begin{abstract}
  We investigate Schwinger pair production of millicharged fermions in
  the strong electric field of cavities used for particle
  accelerators. Even without a direct detection mechanism at hand,
  millicharged particles, if they exist, contribute to the energy loss of the cavity
  and thus leave an imprint on the cavity's quality factor.  Already
  conservative estimates substantially constrain the electric charge of
  these hypothetical particles; the resulting bounds are competitive with the
  currently best laboratory bounds which arise from experiments based
  on polarized laser light propagating in a magnetic field.
  We propose an experimental setup for measuring the electric
  current comprised of the millicharged particles produced in the cavity. 
\end{abstract}

Strong electromagnetic fields offer a new window to particle physics.
Experiments involving strong fields have a new-physics discovery
potential which is partly complementary to accelerator experiments.
For instance, experiments such as BFRT~\cite{Cameron:1993mr},
CAST~\cite{Zioutas:2004hi}, or PVLAS~\cite{Zavattini:2005tm}, using
strong magnetic fields, are involved in the search for light, weakly
coupled particles such as axions.  

Usually, particle physics effects in strong fields result from the 
the macroscopic spatial extent of the
fields which can support coherent phenomena.
If the mass of the new particles is sufficiently low, yet another set
of mechanisms opens up new phenomenological possibilities: processes
can become non-perturbative in the external field, leading to
significant enhancements or increase of phase space.  In a recent
work~\cite{Gies:2006ca}, we have shown that the search for
birefringence and dichroism of polarized laser light propagating in a
strongly magnetized vacuum~\cite{Cameron:1993mr,Zavattini:2005tm}
gives the currently best laboratory bounds on the charge of
millicharged particles. For these bounds, the nonperturbative account
for the magnetic field is crucial.

We would like to stress that improved constraints on millicharged
particles are very welcome. For one thing, the apparently much
stronger astrophysical and cosmological 
bounds~\cite{Dobroliubov:1989mr,Davidson:1991si,% 
  Mohapatra:1990vq,Mohapatra:1991as,Davidson:1993sj,Dubovsky:2003yn}
(for a recent review, see Ref.~\cite{Davidson:2000hf}) have recently
been shown to be quite model dependent~\cite{Masso:2006gc}.  On the
other hand, millicharged particles arise naturally in a large class of
standard model extensions~\cite{Holdom:1985ag,Dienes:1996zr,%
Abel:2003ue,Abel:2004rp,Batell:2005wa}, most notably in a bottom-up 
approach to the string  embedding of the 
standard model~\cite{Aldazabal:2000sa,Abel:2003ue,Abel:2006tbp}.
Therefore, searches for
millicharged particles are a powerful tool to probe fundamental
physics.

The above mentioned laser experiments exploit strong magnetic fields.
One may wonder what can be learned from experiments using strong
electric fields.  There is indeed a paradigm for a nonperturbative
mechanism in strong fields: quantum electrodynamics predicts that
electron-positron pairs are produced from vacuum in strong electric
fields~\cite{Sauter:1931,Heisenberg:1936qt,Schwinger:1951nm}.  A
sizeable rate for spontaneous $e^+e^-$ pair production requires
extraordinary strong electric field strengths $\mathcal E$ of order or
above the critical value ($\hbar=c=1$)
\begin{equation}
{\mathcal E}_{\text{c}}^e \equiv
\frac{m_e^2}{e}
\simeq 1.3\times 10^{18}\ {\rm V/m},
\label{schwinger-crit}
\end{equation}
for which the work of the
field on a unit charge $e$ over the Compton wavelength of the electron,
$\lambdabar =1/m_e$, equals the electron's rest mass $m_e$. 
The process can be viewed as quantum tunneling, giving rise to an
exponential field dependence, $\propto \exp (- \pi \mathcal
E_c^e/\mathcal E)$, which exhibits the nonperturbative structure in
$e\mathcal E$.

Currently, it seems inconceivable to produce macroscopic fields with
electric field strengths of order $\mathcal E \sim
\mathcal{E}_{\text{c}}^e$ in the laboratory\footnote{At the focus of
  standing laser waves, this may eventually be accomplished in the not
  so distant future~\cite{Ringwald:2001ib}.}.  For $\mathcal E\ll
\mathcal E_c^e$, the exponential suppression of the rate makes this
process practically unobservable at present. However, if millicharged
particles with fractional charge $\epsilon = Q_\epsilon/e\ll 1$ and
mass $m_\epsilon$ exist in nature, their corresponding critical
field,
\begin{equation}
{\mathcal E}_{\text c}^\epsilon \equiv
\frac{m_\epsilon^2}{\epsilon e}
\simeq 4.98\times 10^{6}\ \frac{\rm V}{\rm m}\ \frac{1}{\epsilon}
\left( \frac{m_\epsilon}{\rm eV}\right)^2\,,
\label{schwinger-crit-eps}
\end{equation}
may be much smaller and they may be copiously produced with currently
available electric fields.

In this Letter, we want to investigate whether the electric fields
reachable at currently developed accelerator 
cavities will allow for a competitive search for millicharged
particles.  

For a first estimate, we approximate the electromagnetic
field in such a cylindrical cavity as a spatially uniform electric
field, pointing along the cylinder $z$ axis and oscillating with a
frequency $\omega$,
\begin{equation}
{\mathbf E}(t) =(0,0,{\mathcal E}(t))=(0,0,{\mathcal E}_{0}\sin (\omega t))\,,
\hspace{6ex}
{\mathbf B}(t) =(0,0,0)\,.
\label{electric-type}
\end{equation}
For a real cavity, this corresponds to the field configuration on the
$z$ axis.  Typical parameters are ${\mathcal E}_{0}=(35-150)$~MV/m and
$\nu\equiv\omega/2\pi = 1$~GHz, corresponding to $\omega = 4.13\times
10^{-6}$~eV~\cite{Lilje:2004ib,Corsini:2006kb}. Furthermore, we assume
that the frequency $\omega$ is much smaller than the rest energy of
the millicharged particle, $\omega\ll m_\epsilon$.  Under these
conditions, the dominant contribution to the pair-production rate,
i.e., the probability that a pair is produced per unit time and unit
volume, is given by the Schwinger formula \cite{Schwinger:1951nm},
\begin{equation}
\label{schwinger}
w=\frac{{\rm d}^4\,n}{{\rm d}^3x\,{\rm d}t}
=\frac{(2s+1)}{2}\frac{m^{4}_{\epsilon}}{(2\pi)^3}
\left(\frac{{\mathcal{E}}}{\mathcal{E}_{c}^\epsilon}\right)^{2}\sum^{\infty}_{n=1}
\frac{\beta_n}{n^2}\exp\left(-n\pi\frac{\mathcal{E}_{c}^\epsilon}{{\mathcal{E}}}\right),
\end{equation}
where $\beta_n=(-1)^{n+1}$ for bosons and $\beta_n=1$ for fermions;
$s$ denotes the spin of the produced particles \cite{Marinov:1972nx}.
For our quantitative estimates, we will from now on consider fermions,
$s=1/2$, $\beta_n=1$. Corrections to this leading-order formula for
inhomogeneous fields can be computed in a semiclassical manner, using
generalized WKB~\cite{Brezin:1970,Piazza:2004sv}, imaginary-time
methods~\cite{Popov:1971}, propagator constructions
\cite{Dietrich:2003qf}, or modern worldline/instanton methods
\cite{Gies:2005bz,Dunne:2005sx,Kim:2000un}, as well as functional
techniques \cite{Avan:2002dn}. For instance, corrections to the
Schwinger formula for time-like inhomogeneities as in
Eq.~\eqref{electric-type} are controlled by the ratio $\eta$ of the
energy of the laser photons over the work of the field on a charge
$\epsilon e$ over the Compton wavelength of the fermion,
\begin{equation}
\label{gamma}
\eta = \frac{\omega m_{\epsilon}}{\epsilon e {\mathcal E}_{0}}
=
\frac{\omega}{m_\epsilon}\,\frac{{\mathcal E}_c^\epsilon}{\mathcal E_{0}},
\end{equation}
playing the role of an adiabaticity parameter. Incidentally, a similar
parameter exists for spatial inhomogeneities
\cite{Nikishov:1970br,Dunne:2005sx}. Our bounds will, in fact, satisfy the
adiabatic condition, 
\begin{equation}
\eta\ll 1\quad
\Leftrightarrow
\quad
\epsilon\gg 1.4\times 10^{-6}\left(\frac{m_{\epsilon}}{\rm{eV}}\right)
\left(\frac{\nu}{\rm{GHz}}\right)
\left(\frac{50\ \rm{MV/m}}{{\mathcal{E}}_{0}}\right).
\end{equation}
%

%%%%%%%%%%%%%%%%%%%%%%%
\begin{figure}[t]
\begin{center}
\includegraphics[bbllx=48,bblly=330,bburx=574,bbury=602,width=10.5cm]{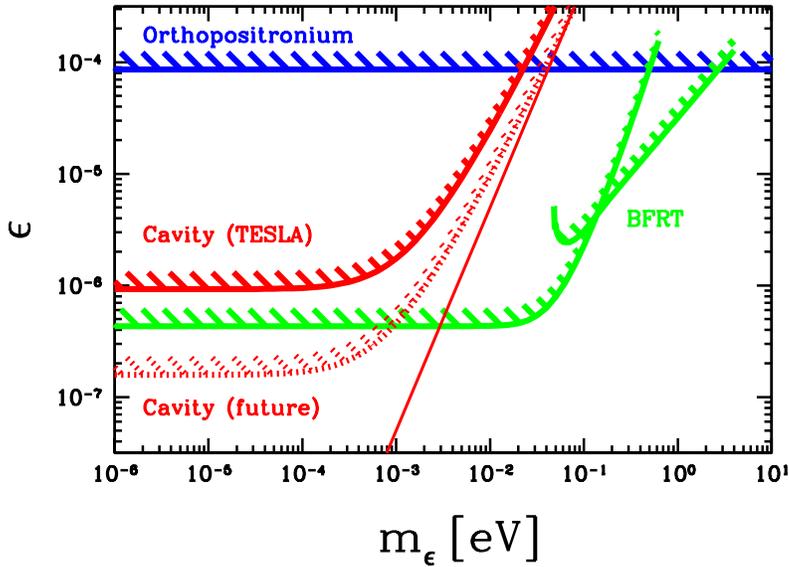}
\vspace{11ex}
\end{center}
\caption[...]{Laboratory limits on the fractional electric charge
  $\epsilon\equiv Q_\epsilon/e$ of a millicharged fermion of mass
  $m_\epsilon$.  The ``Orthopositronium'' limit stems from a limit on
  the branching fraction of invisible orthopositronium
  decay~\cite{Mitsui:1993ha}.  The green ``BFRT'' upper limits
  arise~\cite{Gies:2006ca} from the upper limit on vacuum magnetic
  dichroism and birefringence placed by the laser polarization
  experiment BFRT~\cite{Cameron:1993mr}.  The red (thin solid) line
  corresponds to the (too) naive bound obtained from Eq.
  \eqref{sens_est} (${\mathcal{E}}_{0}=25$~MV/m).  The solid red
  ``Cavity (TESLA)'' upper limit arises from the bound on the energy loss caused by
  Schwinger pair production of millicharged particles in accelerator
  cavities developed for TESLA~\cite{Lilje:2004ib} (${\mathcal{E}}_{0}=25$~MV/m, $L_{\rm{cav}}=10\,\rm{cm}$,
  $Q^{\rm min}_{\rm{MCP}}= 10^{10}$). The red dashed upper limit demonstrates
  the possible bounds obtainable in the near future
  (${\mathcal{E}_0}=50$~MV/m, $L_{\rm{cav}}=10\,\rm{cm}$,
  $Q^{\rm min}_{\rm{MCP}}= 10^{12}$).
\label{fig:sens}}
\end{figure}
%%%%%%%%%%%%%%%%%%%%%%%%

Let us start with an order-of-magnitude estimate of the sensitivity of
accelerator cavities to millicharged fermions.  Assuming that
significant pair production leads to measurable deviations from the
standard electrodynamical behaviour of the cavity, the non-observation
of such deviations implies an upper bound on $\epsilon$ as a function
of $m_\epsilon$.  Using the observation that sizeable pair production
sets in for ${\mathcal E}_{0}/{\mathcal E}_c^\epsilon \sim 0.1-0.25$
\cite{Alkofer:2001ik,Roberts:2002py}, the equation ${\mathcal
  E}_{0}/{\mathcal E}_c^\epsilon = \kappa = {\mathcal O}(0.25)$
translates into 
\begin{equation}
\label{sens_est}
\epsilon \lesssim 2.5\times 10^{-2}
\left( \frac{m_\epsilon}{\rm eV}\right)^2
\left( \frac{\kappa}{0.25}\right)
\left( \frac{50\ {\rm MV/m}}{{\mathcal E}_{0}}\right)
.
\end{equation}
This rough estimate looks very promising. For $m_\epsilon \lesssim
1$~meV, the sensitivity is better than the one obtained from the
observed laboratory limit on vacuum magnetic dichroism due to
pair production of millicharged fermions from laser photons in a
static magnetic field~\cite{Gies:2006ca} (cf. Fig.~\ref{fig:sens}).  Note, that the adiabaticity
parameter, for $\omega=4\times 10^{-6}$~eV, $m_\epsilon \geq 4\times
10^{-4}$~eV, and ${\mathcal E}_0/{\mathcal E}_c^\epsilon = 0.25$, is
indeed small, $\eta \leq 0.04\ll 1$.

For a more realistic estimate of the sensitivity, we have to take into
account that the effects caused by the millicharged particles
typically decrease with shrinking $\epsilon$. Direct detection, for
instance, is therefore not straightforward.  However, 
even without a direct detection mechanism at hand, 
millicharged particles may leave an observable imprint on the properties of
the cavity. 
In particular, 
if a large number of them is produced, they contribute to the macroscopic 
energy loss of the cavity. This 
will be reflected by a decrease of the cavity's quality factor $Q$,
\begin{equation}
\label{qvalue}
Q\equiv 2 \pi E_{\rm cav} / \Delta E ,
\end{equation}
where $E_{\rm cav}$ is the energy stored in the cavity und
$\Delta E$ is the energy loss per oscillation period.
In our case, the latter consists of two parts,
\begin{equation}
  \Delta E = \Delta E_{\rm diss} + \Delta E_{\rm MCP},
\end{equation}
where $\Delta E_{\rm diss}$ is the normal dissipative energy loss in absence
of millicharged particles, and
$\Delta E_{\rm MCP}$ the energy loss into millicharged particles.

In
the following, $\Delta E_{\rm MCP}$ will be estimated by a series of
conservative approximations. For our idealized cavity (cf. Eq.~\ref{electric-type}),  
our resulting $\Delta E_{\rm MCP}$ can hence be viewed as a lower bound 
to the true energy loss.

First, we consider only the kinetic energy carried away by those
millicharged particles which leave the cavity.  At sufficiently high
field strength ${\mathcal{E}}\gtrsim {\mathcal{E}}^{\epsilon}_{c}$,
the particles will predominantly be produced moving highly
relativistic in the direction of the electric field
\cite{Alkofer:2001ik}.  Depending on where and when they are produced,
they may eventually reach a wall of the cavity. For our conservative
estimate, we require that the particle reaches the wall of the cavity
before the direction of the electric field is reversed. For example, a
particle being produced at a time $t$ within the first half
of the oscillation period, $0\leq t\leq \pi/\omega$,
has to reach the wall before $t_{\text r}=\pi/\omega$.  Therefore,
only particles with a distance less than
\begin{equation}
L_{\rm{max}}(t)=\frac{\pi}{\omega}-t
\end{equation}
from the ends of the cylindrical cavity contribute in our estimate. A
particle starting at rest at $t$ and leaving the cavity at
$t_{\text{r}}=\pi/\omega$ has picked up an average energy
\begin{equation}
E_{\rm{av}}(t)=\epsilon e
\frac{1}{L_{\text{max}}(t)}\int_0^{L_{\text{max}}(t)} dL\,
\int_{t}^{t+L} dt'\, \mathcal E(t') =\epsilon e {\mathcal{E}}_{0}
\left(\frac{\cos(\omega t)}{\omega}+\frac{\sin(\omega t)}{\omega
(\pi-t\omega)}\right),
\end{equation}
where the $t'$ integral determines the average field which the
particle is exposed to if it starts at a distance $L$ from the wall.
The $L$ integral is an average over the possible initial positions
$L\leq L_{\rm{max}}$. This implies for the energy loss in one period, 
\begin{equation}
\Delta E_{\rm{MCP}}=4 A_{\rm{cav}}\int^{\pi/\omega}_{0}dt\,
E_{\rm{av}}(t) L_{\rm max}(t)w(t), 
\end{equation}
where $A_{\rm{cav}}$ is the area of the cavity perpendicular to the
electric field. Here, we took a factor two for the second half of the
oscillation period into acount, and another factor of two takes care
of the fact that the energy loss happens at both ends of the cavity.
For the electromagnetic field (\ref{electric-type}), the maximal
energy stored in the cavity is given by
\begin{equation}
E_{\rm{cav}}=\frac{1}{2}{\mathcal{E}}^{2}_{0} A_{\rm{cav}} L_{\rm{cav}},
\end{equation}
with the total length of the cavity $L_{\rm{cav}}$.  The cavity's $Q$
factor is limited by the energy loss into millicharged particles. The
maximal value is reached by an ideal cavity where $\Delta
E_{\rm{diss}}=0$,
\begin{equation}
Q_{\rm{MCP}}=\frac{\pi}{4} \frac{{\mathcal{E}}^{2}_{0}
  L_{\rm{cav}}}{\int^{\pi/\omega}_{0}dt\,  E_{\rm{av}}(t)
  L_{\rm max}(t)w(t)}. 
\end{equation}
Modern superconducting cavities of the type developed for the Tera Electronvolt
Superconducting Linear Accelerator (TESLA) reach $Q$ factors
in excess of $Q_{\rm measured}>10^{10}$ at field strength
${\mathcal{E}}_{0}\sim 25$~MV/m~\cite{Lilje:2004ib}.  Real cavities must have $Q<Q_{\rm
  MCP}$. This enforces $Q_{\rm{measured}}<Q_{\rm{MCP}}$ and constrains
the allowed values of $(\epsilon,m_{\epsilon})$.  The resulting bound
is plotted in Fig.  \ref{fig:sens} and compared to other laboratory
bounds on millicharged particles. Note that, for small masses, the
upper limit for $\epsilon$ becomes independent of the mass and scales as
\begin{equation}
\epsilon \lesssim 10^{-6}\ \left( \frac{10^{10}}{Q^{\rm min}_{\rm MCP}}\right)^{1/3}
\left( \frac{50\ {\rm MV/m}}{{\mathcal E}_0}\right)^{1/3}
\left( \frac{L_{\rm cav}}{30\ {\rm cm}}\right)^{1/3},
\label{sensitivity}
\end{equation}
where $Q^{\rm min}_{\rm MCP}$ is the minimal value for $Q_{\rm MCP}$
allowed by experiment. With the above argument, $Q^{\rm min}_{\rm
  MCP}$ agrees with $Q_{\rm measured}$. 
In the larger mass region, the limit weakens considerably. Improvement in this
region requires stronger electric fields; roughly, the scaling behaviour is
as in Eq.~(\ref{sens_est}). 

An even better bound could be obtained by comparing the expected $Q$
factor, $Q_{\rm{expected}}$, for a real cavity in absence of
millicharged particles with the measured $Q$ factor, 
$Q_{\rm{measured}}$. If millicharged particles exist, the measured $Q$
factor should deviate from the expected one by,
\begin{equation}
\left( Q_{\rm{measured}}^{-1}-Q_{\rm{expected}}^{-1}\right)^{-1}=Q_{\rm{MCP}}.
\end{equation}
A $10\%$ accuracy of the quantities contributing to the left-hand side
gives already an improvement by a factor of $10$ for the lower
limit $Q^{\rm min}_{\rm MCP}$ on $Q_{\text{MCP}}$. Using this and some 
further improvements of cavities, including an 
increase in the maximal field strength by a factor of two, we plot the corresponding 
bound in Fig. \ref{fig:sens} (red dashed line) to demonstrate the potential
sensitivity reachable in the near future. The bound is well
competitive with the one coming from laser experiments which currently
provide the best laboratory bounds.

Further improvement may come from a more accurate determination of the
energy loss caused by millicharged particles. Our criterion for
particles to leave the cavity presumably takes into account only a
fraction of particles that ultimately leave the cavity. On the other
hand, in a real cavity, an additional complication arises due to the
non-vanishing magnetic fields. These will lead to more complicated
trajectories than the ones used in our simple estimate. Nevertheless,
depending on the precise field distribution in the cavity, we expect
that only exceptional particle trajectories inside the cavity are
stable. Since the pairs are generally created with a continuous
momentum distribution, the probability that a particle moves precisely
on a stable trajectory inside the cavity is presumably very small.
Therefore, it may well be that a large fraction of all produced
particles ultimately leaves the cavity; this fraction would thus
contribute to $\Delta E_{\text{MCP}}$ rather independently of the
initial position.  Moreover, the set of produced particles on stable
trajectories will undergo plasma
oscillations~\cite{Kluger:1991ib,Alkofer:2001ik,Roberts:2002py,Ruffini:2003cr} and potentially
contribute indirectly to energy loss.  Finally, it has to be checked
for a real cavity whether some produced pairs could evade to
contribute to the energy loss by subsequent coherent pair
annihilation; for the special case of a spatially homogeneous field,
this effect can lead to a reduction of pair
accumulation~\cite{Roberts:2002py} in comparison with the Schwinger
formula \eqref{schwinger}.  Future estimates may also include the
particles' rest mass which we have not added to the energy loss so
far. Also, thermal fluctuations can lead to an enhancement of the
pair-production rate \cite{Gies:1999vb}.  In total, all these
considerations may well lead to a factor $0.1-100$ in the energy loss
compared to our conservative estimate. Since, for small masses, the
sensitivity in $\epsilon$ scales with $\Delta E_{\rm MCP}^{-1/3}$,
this leads only to a moderate change of the bound by a factor of
$2-0.2$.

Ultimately, one would like to probe also the region of larger masses.
This requires much stronger electric fields.  For example, one may
envisage an experiment which exploits the electric field in an
antinode of a standing wave produced by a superposition of two
petawatt laser beams, focussed to the diffraction limit
(cf.~\cite{Ringwald:2001ib,Heinzl:2006xc}). In such an experiment,
field strengths in excess of $10^{8}\,\rm{MV}/\rm{m}$ may be reached,
and consequently particles with larger masses may be produced.
However, measuring the energy loss into millicharged particles is not
straightforward in this set-up. Nevertheless, this possibility should
be seriously studied, because it would be capable of testing the
millicharged-fermion interpretation of the PVLAS dichroism signal,
which requires $\epsilon\sim 3\times 10^{-6}$ and $m_\epsilon\sim
0.1$~eV~\cite{Gies:2006ca}.

%%%%%%%%%%%%%%%%%%%%%%%
\begin{figure}[t]
\begin{center}
\includegraphics[bbllx=136,bblly=94,bburx=339,bbury=673,angle=-90,width=8.5cm]{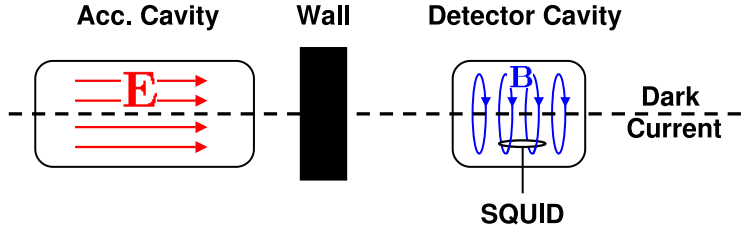}
\vspace{-3ex}
\end{center}
\caption[...]{Schematic set up for a ``dark current shining through a wall'' 
experiment. The alternating dark current (frequency $\nu$), comprised of the produced millicharged particles 
(dashed line), escapes from the accelerator cavity  and traverses also a 
thick shielding (``wall''), in which the conventional dark current of electrons
is stopped. The dark current induces a magnetic field in a resonant (frequency $\nu$) detector cavity  
behind the wall, which is detected by a SQUID~\cite{Vodel:2005ma}.    
\label{fig:detector}}
\end{figure}
%%%%%%%%%%%%%%%%%%%%%%%%

Above, we have discussed how one can obtain bounds on millicharged particles
from the regular operation of accelerator cavities. A more direct approach
to infer the existence of such particles may be based on the detection
of the electrical current comprised of them. 
In Fig.~\ref{fig:detector}, 
we show schematically how one could set up an experiment to detect this current.

{\it In summary:} Schwinger pair production in strong electric fields
could turn accelerator cavities into factories for light millicharged
particles, whose possible existence is, in many extensions of the 
standard model, directly tied to physics at very large energy scales, 
even up to the Planck scale, $M_{\rm P}\sim 10^{19}$~GeV. 
Hence, parts of accelerators may probe higher energy scales than the 
accelerator beams themselves.  

\acknowledgments

We would like to thank  
E.-A. Knabbe, L. Lilje, A. Lindner, P. Schm\"user, and H. Weise
for valuable information on accelerator cavities. 
HG acknowledges support by DFG Gi 328/1-3.

\end{document}